\documentclass[sigconf]{acmart}

\AtBeginDocument{
  }

\setcopyright{acmlicensed}
\copyrightyear{2026}
\acmYear{2026}
\acmDOI{XXXXXXX.XXXXXXX}
\acmConference[Conference acronym 'XX]{Make sure to enter the correct
  conference title from your rights confirmation email}{June 03--05,
  2018}{Woodstock, NY}
\acmISBN{978-1-4503-XXXX-X/2018/06}

\begin{document}

\title{From Hidden Profiles to Governable Personalization: Recommender Systems in the Age of LLM Agents}

\author{Jiahao Liu}
\affiliation{
  \institution{Fudan University}
  \city{Shanghai}
  \country{China}
}
\email{jiahaoliu23@m.fudan.edu.cn}

\author{Mingzhe Han}
\affiliation{
  \institution{Fudan University}
  \city{Shanghai}
  \country{China}
}
\email{mzhan22@m.fudan.edu.cn}

\author{Guanming Liu}
\affiliation{
  \institution{Fudan University}
  \city{Shanghai}
  \country{China}
}
\email{gmliu24@m.fudan.edu.cn}

\author{Weihang Wang}
\affiliation{
  \institution{Fudan University}
  \city{Shanghai}
  \country{China}
}
\email{kiren.wwh@gmail.com}

\author{Dongsheng Li}
\affiliation{
  \institution{Microsoft Research Asia}
  \city{Shanghai}
  \country{China}
}
\email{dongshengli@fudan.edu.cn}

\author{Hansu Gu}
\affiliation{
  \institution{Independent}
  \city{Seattle}
  \country{United States}
}
\email{hansug@acm.org}

\author{Peng Zhang}
\affiliation{
  \institution{Fudan University}
  \city{Shanghai}
  \country{China}
}
\email{zhangpeng\_@fudan.edu.cn}

\author{Tun Lu}
\affiliation{
  \institution{Fudan University}
  \city{Shanghai}
  \country{China}
}
\email{lutun@fudan.edu.cn}

\author{Ning Gu}
\affiliation{
  \institution{Fudan University}
  \city{Shanghai}
  \country{China}
}
\email{ninggu@fudan.edu.cn}

\renewcommand{\shortauthors}{Jiahao Liu et al.}

\begin{abstract}
Personalization has traditionally depended on platform-specific user models that are optimized for prediction but remain largely inaccessible to the people they describe. As LLM-based assistants increasingly mediate search, shopping, travel, and content access, this arrangement may be giving way to a new personalization stack in which user representation is no longer confined to isolated platforms. In this paper, we argue that the key issue is not simply that large language models can enhance recommendation quality, but that they reconfigure where and how user representations are produced, exposed, and acted upon. We propose a shift from hidden platform profiling toward governable personalization, where user representations may become more inspectable, revisable, portable, and consequential across services. Building on this view, we identify five research fronts for recommender systems: transparent yet privacy-preserving user modeling, intent translation and alignment, cross-domain representation and memory design, trustworthy commercialization in assistant-mediated environments, and operational mechanisms for ownership, access, and accountability. We position these not as isolated technical challenges, but as interconnected design problems created by the emergence of LLM agents as intermediaries between users and digital platforms. We argue that the future of recommender systems will depend not only on better inference, but on building personalization systems that users can meaningfully understand, shape, and govern.
\end{abstract}



\begin{teaserfigure}
  \centering
  \includegraphics[width=0.8\textwidth]{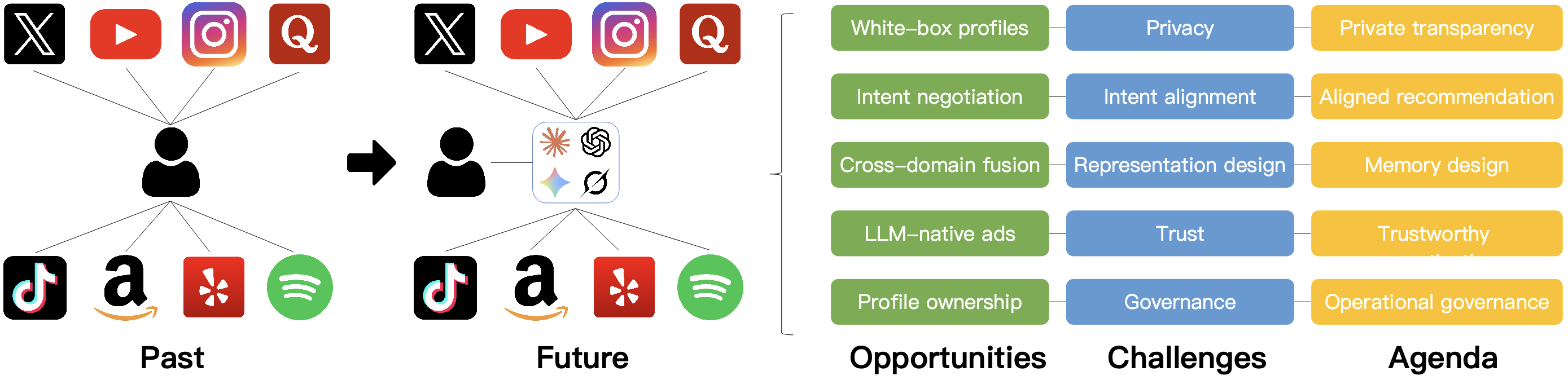}
  \caption{From platform profiles to user-controlled intent layers in LLM-mediated personalization.}
\end{teaserfigure}

\maketitle

\section{Introduction}

The history of recommender systems is, to a large extent, the history of increasingly sophisticated user modeling. For decades, personalization has been organized around a platform-centric architecture: users interact with many services~\cite{adomavicius2005toward}, each service collects behavioral data, and each service constructs its own internal representation of the same individual. In practice, the same person may be represented differently across content platforms, e-commerce services, travel platforms, professional networks, and social media. These representations are inferred locally, stored privately, and optimized toward distinct platform objectives~\cite{koren2009matrix}.

This architecture has created substantial economic and social value. It supports retrieval, ranking, matching, and targeting at scale under conditions of sparsity, latency constraints, and partial observability. Yet it has also institutionalized a basic asymmetry: platforms interpret users, while users rarely understand how they are being interpreted. Conventional user profiles are not designed as user-facing objects. They are machine-oriented artifacts embedded in feature stores, embeddings, sequence models, retrieval indexes, or neural parameters~\cite{hidasi2015session}~\cite{xia2022fire,liu2022parameter,liu2023personalized,liu2023autoseqrec,feng2026drift}, built for prediction and optimization rather than user understanding, contestability, or control. Personalization has thus become highly effective while remaining fragmented~\cite{adomavicius2005toward}, opaque, and platform-owned~\cite{koren2009matrix}.

Large language models (LLMs) suggest that this arrangement may be entering a new phase. LLMs are increasingly becoming entry points to everyday digital activities, including search, shopping, travel planning, scheduling, and content discovery. More importantly, users can now express not only preferences~\cite{jannach2021survey}, but also constraints, exclusions, temporary states~\cite{lin2025can}, and contextual trade-offs in natural language. Rather than leaving systems to infer intent solely from clicks, skips, or dwell time, users can directly specify how they wish to be served. A user may state that they do not want emotionally intense content this week, that they prefer a quiet hotel near a venue rather than an ``Instagrammable'' one, or that they want recommendations without overt commercial pressure. These inputs are not merely additional features; they are explicit articulations of intent~\cite{zhao2024recommender,xu2024prompting}~\cite{liu2025filtering,liu2025improving,wu2025bidirectional}.

The significance of this shift is architectural, not merely interface-level. The emerging ecosystem may no longer be one in which users separately manage many personalized platforms in parallel. Instead, users may increasingly interact with an LLM-based assistant that mediates intent~\cite{yao2022react} and coordinates with multiple downstream services through APIs, plugins, or tools~\cite{schick2023toolformer}~\cite{liu2025agentcf++}. Under these conditions, the central question is no longer only how each platform infers preferences from local behavior. It is also how an assistant represents the user, how that representation is governed, how it is translated into cross-platform action, and how trust is maintained when recommendation, assistance, and commercialization converge within the same conversational layer~\cite{zhao2024recommender}.

This paper advances a simple position: the key transformation is a redistribution of representational control. User modeling is shifting from \textbf{platform-owned black-box profiles} toward more \textbf{user-addressable, partially white-box, and potentially portable intent representations}. This shift does not eliminate platform profiles~\cite{zhang2020explainable}; rather, it introduces an upstream layer that may increasingly mediate them, such that the user profile is no longer only a hidden optimization variable~\cite{jannach2016user} but also a possible interface for user governance~\cite{zhao2024recommender}.

We develop this argument as a position paper rather than as an empirical claim about a single settled architecture. Our goal is to clarify the conceptual shift underway, articulate the main opportunities and risks, and organize a research agenda for the recommender systems community. To do so, we structure the paper around five paired opportunities and challenges: \textbf{profile white-boxization and privacy}, \textbf{profile-mediated negotiation and intent alignment}, \textbf{cross-domain fusion and representation design}, \textbf{LLM-native advertising and trust}, and \textbf{profile ownership and governance}. Together, these dimensions frame both our diagnosis of the emerging ecosystem and the research directions that follow from it.

Accordingly, this paper makes three contributions to frame this discussion. First, it conceptually distinguishes among \textbf{platform profiles}, \textbf{LLM profiles}, and \textbf{user-controlled intent layers}. Second, it develops a structured opportunity and challenge framework for understanding the transition to LLM-mediated personalization. Third, it outlines a forward-looking research agenda for recommender systems organized around the same five dimensions.

\section{A Conceptual Framework for User-Controlled Intent Layers}

A clearer conceptual vocabulary is necessary if LLM-mediated personalization is to be analyzed precisely. In particular, it is important to distinguish among \textbf{platform profiles}, \textbf{LLM profiles}, and \textbf{user-controlled intent layers}, because these forms of representation differ not only in technical form, but also in visibility, editability, and governance.

A \textbf{platform profile} is fundamentally an optimization artifact. It is constructed to improve matching, ranking, retention, and conversion~\cite{adomavicius2005toward}, and is typically encoded in embeddings, feature interactions, sequence models~\cite{hidasi2015session}, or neural parameters. Such representations are computationally effective, but they are usually black-box~\cite{koren2009matrix}, difficult to interpret, and difficult to revise directly. Even when platforms provide controls such as ``less of this,'' ``hide ad,'' or history deletion, these remain indirect interventions on an underlying model that is largely inaccessible to the user~\cite{jannach2016user}.

An \textbf{LLM profile} differs in an important respect. It may incorporate explicitly stated identity and preference information, for example, ``My name is Kimi,'' ``I am a recommender-systems engineer,'' or ``I prefer technical explanations and do not want high-sugar food recommendations by default,'' while also summarizing repeated interactions into persistent memory~\cite{packer2023memgpt}. Unlike conventional platform profiles, such representations can often be surfaced in natural language~\cite{zhang2025survey}, inspected, corrected, and refined through dialogue. Their importance lies not only in increased visibility, but also in the user's more direct participation in shaping the representation itself~\cite{park2023generative}.

Even so, the notion of a \emph{profile} remains too narrow for the emerging personalization setting. A profile implies a relatively stable description of a person. What users increasingly require, however, is not simply a description of who they are, but a means of specifying how they wish to be represented and served under changing conditions. For this reason, we propose the broader concept of a \textbf{user-controlled intent layer}. An intent layer includes relatively stable identity information, but also temporary goals, short-term exclusions, contextual preferences, scenario-specific trade-offs, tolerance thresholds, and authorization boundaries. It does not merely record who the user has been; it helps govern how the assistant should interpret and act for the user in the present context.

This distinction matters because human preferences are plural, contextual, and dynamic. The same person may prefer efficiency in business travel but aesthetics in leisure travel, strict filtering during stressful periods but exploration at other times, and different recommendation styles in professional and social settings. Under the traditional paradigm, the profile is primarily the product of a platform's inference~\cite{pu2008user}. Under the emerging paradigm, representation increasingly becomes an interface through which users articulate policies~\cite{jannach2021survey}, boundaries, and priorities that shape downstream recommendation and action~\cite{he2016interactive}.

Making profiles user-addressable does not reduce complexity; it relocates it. Once representations become visible and revisable, new questions emerge. Which parts of the representation should be treated as stable and which as contextual? Which inferences may be generated automatically, and which require explicit consent? How should systems reconcile conflicts among stated intent, observed behavior, inferred long-term interests, and downstream platform objectives? The move toward user-controlled intent layers therefore does not simplify recommendation; it shifts the challenge from hidden model fitting alone to the explicit design and governance of representations~\cite{jannach2016user,he2016interactive}.

\section{Opportunities and Challenges}

\paragraph{White-boxization and privacy.}
A first opportunity is \textbf{profile white-boxization}. In conventional recommender systems, users often sense that a platform has formed some interpretation of them, but rarely know what that interpretation is, how it was formed, or how far it has drifted from their actual intentions. An LLM can help narrow this gap by observing platform outputs, user reactions, and cross-session instructions, and by surfacing not only recommendations, but also the assumptions underlying them. In this sense, personalization may become more reflexive: the system can expose how a user is being represented~\cite{tintarev2007survey}, where overgeneralization has occurred, and when temporary behavior has been misread as durable preference~\cite{zhang2020explainable}~\cite{liu2023triple}.

The corresponding challenge is \textbf{privacy under white-box and inferential profiling}. A richer and more legible profile may empower users, but it may also create a more comprehensive and potentially more invasive representation of their lives. An LLM that observes multiple contexts may infer latent attributes never explicitly disclosed~\cite{himeur2022latest}, reconstruct how other platforms already perceive the user~\cite{fan2022comprehensive}, or aggregate sensitive information across domains in ways no single platform previously could. The privacy problem is therefore not only about storage or access control, but also about inferential exposure: what may be derived, who may access those derivations, and which inferences should remain unavailable unless explicitly authorized~\cite{regulation2016regulation}~\cite{han2025fedcia,han2026feature}.

\paragraph{Negotiation and intent alignment.}
A second opportunity is that user-addressable profiles can become a \textbf{negotiation language between users and platforms}. Historically, users had only weak mechanisms for reshaping recommendation~\cite{pu2008user}, such as creating new accounts, deleting histories, or repeatedly signaling ``not interested.'' With an LLM-mediated layer, users may specify constraints more directly~\cite{jannach2016user}: they may request temporary suppression of a topic, broader exploration without sensationalism, or shopping suggestions that prioritize durability over prestige. In this way, the profile becomes a lever through which users can intervene in the recommendation loop itself rather than merely reacting to its outputs~\cite{jannach2021survey}.

The corresponding challenge is \textbf{intent alignment}. Once users can specify what should be recommended, filtered, or prioritized, the assistant must interpret, reconcile, and operationalize those instructions. Yet users do not possess perfectly stable or internally unified preferences: they have short-term desires and long-term goals, temporary moods and durable values, appetites and self-protective constraints. The central question is therefore which version of the user the system ought to serve. Should it optimize for immediate attraction, explicit instruction, inferred well-being, or consistency with prior commitments? This is not merely a technical problem of intent parsing, but a normative problem concerning autonomy~\cite{jesse2021digital}, paternalism, and legitimate forms of intervention~\cite{fan2022comprehensive}.

\paragraph{Cross-domain fusion and representation.}
A third opportunity concerns \textbf{cross-domain interest fusion}. Many real needs are inherently cross-domain: a career transition may shape what a person wants to read, learn, buy, and whom they wish to connect with; a health goal may affect food ordering, exercise planning, sleep guidance, and shopping behavior; a travel decision may depend simultaneously on schedule, budget, mobility constraints, social context, and aesthetic preference. Historically, such integration has been limited because platforms see only fragments of the user and because no shared, user-understandable representation of intent has existed across domains. LLMs make such fusion more plausible by operating over meaning~\cite{chen2024survey} and organizing disparate signals into a more coherent narrative, schema, or policy set~\cite{lin2025can,wu2024survey}.

The corresponding challenge is \textbf{representation and memory design}. To support personalization, the profile must be interpretable, editable, fine-grained, and computationally useful at once. No single representational form fully satisfies these requirements. Natural language is expressive and user-readable, but unstable and difficult to compute over reliably. Embeddings support retrieval and generalization, but provide weak affordances for direct control and contestability. Flat tags are legible, but often too coarse to capture conditionals, exceptions, and temporal scope. A promising direction is therefore a layered architecture that combines natural language for expression and explanation~\cite{packer2023memgpt}, structured semantic representations for scope, time, confidence, provenance, and policy constraints~\cite{zhang2025survey}, and latent representations for retrieval, matching, and generalization~\cite{park2023generative}. Memory design is central: profiles should support \textbf{progressive disclosure}, such that inferred elements differ in visibility, durability, and actionability~\cite{zhao2024recommender}.

\paragraph{LLM-native advertising and trust.}
A fourth opportunity is \textbf{LLM-native advertising}. As assistants become entry points to shopping, travel, ride-hailing, food delivery, and other services, external platforms and merchants may increasingly expose callable tools that allow assistants to compare, recommend, and transact across services directly. This creates a new commercial environment in which paid products and services may be embedded within the assistant experience itself. In domains where assistant replies surface only one or two options, ranking position may become even more consequential than in traditional list-based interfaces~\cite{edelman2007strategic}~\cite{liu2026distribution,tong2026rq}. The assistant layer may therefore emerge as a new locus of sponsorship, bidding, and marketplace power~\cite{zhao2024recommender}.

The corresponding challenge is \textbf{trust and commercial design}. Search and recommendation have often been experienced as user-serving services, while advertising has remained a partially separable commercial layer. In assistant-mediated interaction, that separation becomes harder to sustain. When the assistant speaks in the voice of advice, commercial influence may appear less like exposure and more like manipulation~\cite{jesse2021digital}. The design challenge is therefore how assistants can monetize without eroding the trust required for delegation. Users are more likely to accept commercial content only if sponsorship is legible~\cite{deng2016reducing}, promotion does not override user-specified constraints~\cite{wilkinson2021pursuit}, the boundary between eligibility and paid preference remains intelligible, and enforceable norms exist around disclosure, fairness, and recourse~\cite{fan2022comprehensive,chiarella2023digital}.

\paragraph{Profiles as digital assets and governance.}
A fifth opportunity is to treat profiles as \textbf{personal digital assets}. At the firm level, this may reduce cold-start costs and weaken data lock-in by allowing services to operate on user-authorized representations rather than reconstructing isolated models from scratch~\cite{mansour2016demonstration}. At the user level, such profiles may become useful across collaboration, professional representation, social discovery, and personalized service orchestration. They may function less as static dossiers than as dynamic, permissioned personal cards selectively shared for specific purposes~\cite{regulation2016regulation}.

The corresponding challenge is \textbf{governance}. Portability and ownership sound empowering, but they remain shallow if users lack meaningful control over interpretation, sharing, retention, and reuse. A portable profile without revocation rights~\cite{mansour2016demonstration}, selective disclosure, purpose limitation, or auditability may simply reproduce the asymmetries of platform-centric personalization in a new form. The key issue is therefore not only whether profiles can move across services, but under what rules: how users grant scoped access, how downstream actors are constrained in secondary use, how sensitive inferences are protected from surveillance and discrimination~\cite{regulation2016regulation}, and how users may contest, withdraw, or narrow access over time~\cite{chiarella2023digital}.

\section{A Research Agenda for the Recommender Systems Community}

The five paired opportunities and challenges above do more than characterize the transition to LLM-mediated personalization; they also define the main research directions that follow from it. Consistent with the position paper orientation of this work, we present these directions as a structured agenda for the field rather than as a finalized taxonomy. We organize the agenda below along the same five dimensions.

\paragraph{From white-boxization to privacy-preserving profile transparency.}
If profile white-boxization is one of the defining opportunities of LLM-mediated personalization, then a first research priority is to determine how profiles can become more legible without becoming more intrusive. Recommender systems research should move beyond improving inference alone toward \textbf{representation governance}: systems should enable users to inspect, revise, contest, and govern the representations through which they are served. This requires formalizing properties such as legibility, editability, provenance, reversibility, and permissioning, while also developing mechanisms for limiting inferential overreach, constraining sensitive attribute discovery, and determining which inferred elements should be visible, suppressible, or inaccessible by default~\cite{liu2023recommendation}. The central challenge is therefore not transparency per se, but \textbf{privacy-preserving transparency}.

\paragraph{From profile-mediated negotiation to intent-aligned recommendation.}
If user-addressable profiles become a negotiation language between users and platforms, then recommender systems must be able to translate user instructions into reliable downstream behavior. This motivates research on \textbf{layered intent architectures} that connect natural-language self-description, structured policy representations, and latent computational forms. Such architectures must support the propagation of user-specified constraints into retrieval, ranking, generation, and action, while also handling intent parsing, policy translation, memory consolidation, and conflict resolution. These issues become especially difficult when short-term desires, long-term goals, prior commitments, and inferred interests diverge. The key challenge is not merely whether systems can interpret user intent, but whether they can remain \textbf{aligned} with the version of the user they are expected to serve.

\paragraph{From cross-domain fusion to robust representation and memory design.}
If LLMs enable cross-domain fusion, then the field must address how integrated user representations should be structured, stored, updated, and operationalized. This makes \textbf{representation and memory design} a central research agenda. Work is needed on architectures that combine natural language for user expression and explanation, structured semantic layers for scope, time, confidence, provenance, and policy constraints, and latent representations for retrieval and generalization. Systems should also support \textbf{progressive disclosure}, so that inferred elements differ in visibility, durability, and actionability rather than being treated uniformly. Closely related is the design of \textbf{interfaces for memory, scope, and disclosure}: if profiles become user-addressable, users will need ways to inspect what is known, distinguish what is stated from what is inferred, correct errors, and manage the duration, scope, and visibility of profile elements across contexts.

\paragraph{From LLM-native advertising to trustworthy commercial architectures.}
If assistants become upstream gatekeepers to downstream services, then the economics of recommendation change as well. This creates a research agenda around \textbf{LLM-native advertising} and the preservation of trust under commercialization. Recommender systems research should revisit marketplace design, sponsored ranking, incentive alignment, and allocation fairness in agent-mediated environments~\cite{liu2025unbiased,zhang2025evalagent,gu2026llm}. Questions that were once secondary may become central: how sponsored candidates should be integrated into assistant outputs, how conflicts of interest should be disclosed, how eligibility and paid promotion should be separated, and how user-specified constraints should be protected from commercial override. The central challenge is to design monetization mechanisms that remain \textbf{compatible with trust and delegation}.

\paragraph{From profile ownership to operational governance.}
If profiles are to function as personal digital assets, then ownership cannot be reduced to portability alone. The field therefore needs a research agenda on \textbf{governance as an internal design problem of recommender systems}. This includes selective disclosure, scoped authorization, revocation, purpose limitation, auditability, appeal mechanisms, sponsorship disclosure, and anti-discrimination safeguards. Governance should be implemented not as an external afterthought, but within the representational and interactional fabric of the system itself: in schemas, defaults, interfaces, permissions, audit trails, and optimization objectives. A portable profile becomes meaningful only if users can also control how it is interpreted, reused, and constrained over time.

\section{Conclusion}

In this position paper, we argued that recommender systems may be entering a transition from platform-centric personalization to LLM-mediated personalization. Under the earlier paradigm, user profiles were hidden optimization artifacts: powerful, predictive, and controlled by platforms. Under the emerging paradigm, user representations may become explicit, editable, portable, and consequential in shaping how people access digital services. We therefore argued that the meaning of the profile is changing: it becomes not only a description of the user, but also a interface through which the user governs recommendation, filtering, and delegation. This shift creates opportunities, profile white-boxization, profile-mediated negotiation, cross-domain fusion, LLM-native advertising, and profile ownership as a personal digital asset, but also challenges in privacy, alignment, representation, trust, and governance. For the recommender systems community, the central implication is that the next frontier may depend less on making systems better at inferring users from traces, and more on making users better able to shape how they are represented and served across contexts. In this sense, the future of recommendation is not only about better prediction or better ranking; it is also about designing \textbf{user-controlled intent layers} that make personalization more controllable, interpretable, portable, and trustworthy in the age of LLM-mediated interaction.

\balance

\bibliographystyle{ACM-Reference-Format}
\bibliography{main}

\end{document}